\documentclass[aps,reprint,amsmath,amssymb]{revtex4-2}

\usepackage{graphicx}
\usepackage{dcolumn}
\usepackage{bm}
\usepackage[version=4]{mhchem}
\usepackage{amsmath}
\usepackage[mathlines]{lineno}

\newcommand{\SiVmm}{SiV\textsuperscript{2$-$}}
\newcommand{\SiVm}{SiV\textsuperscript{$-$}}
\newcommand{\SiVo}{SiV\textsuperscript{0}}
\newcommand{\SiVp}{SiV\textsuperscript{+}}

\renewcommand{\deg}{$^{\circ}$}
\newcommand{\figref}[1]{Figure \ref{#1}}

\begin{document}

\title{Solution gate control of shallow silicon vacancy charge states in diamond}

\author{Charlie Pattinson\textsuperscript{1}}
\email{charlie.pattinson@unimelb.edu.au}
\author{Daniel J. McCloskey\textsuperscript{1}}%
\author{Nikolai Dontschuk\textsuperscript{1}}%
\author{Brett C. Johnson\textsuperscript{2}}
\author{Alexander A. Wood\textsuperscript{1}}
\author{David Simpson\textsuperscript{1}}%
 \email{simd@unimelb.edu.au}
\affiliation{%
  \textsuperscript{1}School of Physics, The University of Melbourne, Victoria 3010, Australia
}%
\affiliation{%
\textsuperscript{2}School of Science RMIT University Melbourne, VIC 3001, Australia
}

\date{\today}

\begin{abstract}
Silicon-vacancy (SiV) centers in diamond combine near-infrared emission with solid-state robustness, but their performance hinges on isolating favorable defect charge states. We demonstrate static and dynamic control of ultra-shallow ($<$15 nm) SiV ensembles in type IIa diamond. By combining low-energy ion implantation with tailored oxygen and hydrogen terminations, we map regimes that maximise the fluorescent \SiVm population over dark charge states. We then realize reversible \SiVm $\leftrightarrow$ \SiVo conversion using aqueous electrolytic gating with sub-200 mV biases and low optical powers. Our results enable low-power electrical control of SiV ensembles for integrated quantum photonics and biologically compatible voltage imaging in the near-infrared. 
\end{abstract}

\maketitle

Point defects in diamond have emerged as leading solid-state platforms for quantum technologies, enabling single-photon emission, spin-based quantum information processing and nanoscale sensing and imaging under ambient conditions \cite{kurtsieferStableSolidStateSource2000,beveratosSinglePhotonQuantum2002,sipahigilIndistinguishablePhotonsSeparated2014,waltrichTwophotonInterferenceSiliconvacancy2023,mccloskeyDiamondVoltageImaging2022}. Among these, the nitrogen-vacancy (NV) center has been extensively explored owing to its long spin coherence time and convenient optical spin readout \cite{bar-gillSolidstateElectronicSpin2013,gruberScanningConfocalOptical1997a}. However, the NV center suffers from a broad phonon sideband, a relatively low Debye–Waller factor and strong spectral diffusion near surfaces, which pose challenges for integration in photonic devices and for spectrally multiplexed biological imaging environments.
The negatively charged silicon-vacancy (\SiVm) center has recently attracted significant interest as a complementary defect with highly favorable optical properties. Owing to its inversion-symmetric structure, the \SiVm center exhibits a narrow zero-phonon line (ZPL) at 737 nm with a large Debye–Waller factor ($\sim$0.70) \cite{neuSinglePhotonEmission2011}, making it particularly promising for integrated quantum photonic devices, high-brightness single photon sources and biological imaging \cite{liuSiliconVacancyNanodiamondsHigh2022,golubewaAllOpticalThermometryNV2022}. SiV centers are known to exist in multiple charge states (\SiVmm, \SiVm, \SiVo), however \SiVm\, is the only charge state known to fluoresce in the near-infrared at room temperature\cite{dhaenens-johanssonOpticalPropertiesNeutral2011}. The relative populations of these charge states are governed by the Fermi level, the local defect environment and band bending at the diamond surface \cite{roseObservationEnvironmentallyInsensitive2018,zhangNeutralSiliconVacancy2023}. Consequently, both static and dynamic control of the SiV charge state are essential for high-fidelity operation of SiV-based devices and for the development of charge-state-based sensing modalities.
For individual near-surface SiV centers, it has been shown that surface chemistry and band bending can modify charge-state occupancy and fluorescence\cite{zhangNeutralSiliconVacancy2023,pasternakEffectHTerminatedSurfaces2025}. However, systematic charge-state control of shallow SiV ensembles, particularly using a combination of depth engineering, surface termination and in-solution gating, remains comparatively unexplored. This gap is technologically relevant as many emerging applications in sensing and imaging will rely on shallow ensembles rather than isolated emitters. Ensembles provide enhanced signal levels and improved robustness, whereas individual defects within an ensemble experience substantially different local electrostatic and strain environments \cite{angellUnravelingSourcesEmission2024,zuberShallowSiliconVacancy2023}, making the ensemble charge-state response inherently inhomogeneous and more difficult to control.
In this work, we investigate static and dynamic charge-state control of shallow SiV ensembles implanted in Type IIa single-crystal diamond. By combining low-energy silicon ion implantation with controlled oxygen and hydrogen surface terminations, we elucidate the interplay between SiV depth, surface-induced band bending and the resulting charge-state populations. Using photoluminescence spectroscopy, we show that surface hydrogenation increases the near-surface \SiVm ensemble population at the expense of the non-fluorescent \SiVmm charge state, and that the implantation depth is a key parameter for optimising the fluorescent \SiVm fraction. Building on this engineered initial state, we demonstrate dynamic, low-voltage electrical manipulation of the SiV charge state using an aqueous electrolytic (solution) gate applied to the hydrogen-terminated diamond surface. Under both green (532 nm) and red (660 nm) excitation, we observe monotonic, reversible changes in SiV fluorescence with applied gate bias, consistent with electrically assisted interconversion between \SiVm and \SiVo. The achieved contrasts and operating voltages compare favourably with previous NV-based charge-control schemes \cite{mccloskeyDiamondVoltageImaging2022}. Our results establish a pathway towards high-fidelity, compact and low-power electrical control of SiV charge state in shallow ensembles, and highlight their potential as a platform for integrated quantum photonic devices and in-solution voltage sensing.

\begin{figure}[!h]
    \includegraphics{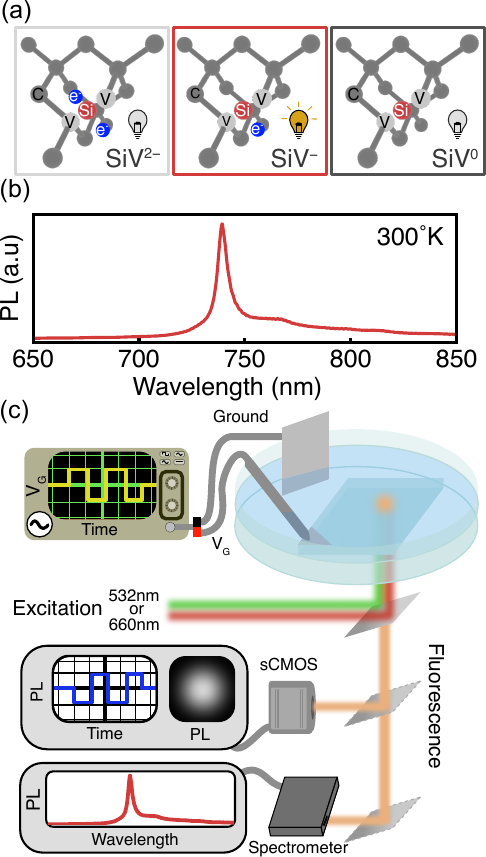}
    \caption{Optical characteristics of silicon vacancy center charge states in diamond and solution gating set up.(a) Schematic of the three distinct silicon vacancy charge states in diamond and (b) the photoluminescence spectrum of the only room temperature fluorescent charge state \SiVm. (c) Schematic of the solution-gating apparatus used to characterize the voltage response. The voltage from a signal generator is applied to the diamond via a platinum electrode in contact with a Ti/Pt metal contact deposited on the corner of the diamond surface. The sample is illuminated via either 532nm or 660nm excitation using a widefield inverted fluorescence microscope. The \SiVm fluorescence is collected using a 700nm long pass dichroic onto a sensitive sCMOS camera or a spectrometer via a flip mirror. }
    \label{fig: figure1}
\end{figure}

\begin{figure*}[t]
    \includegraphics{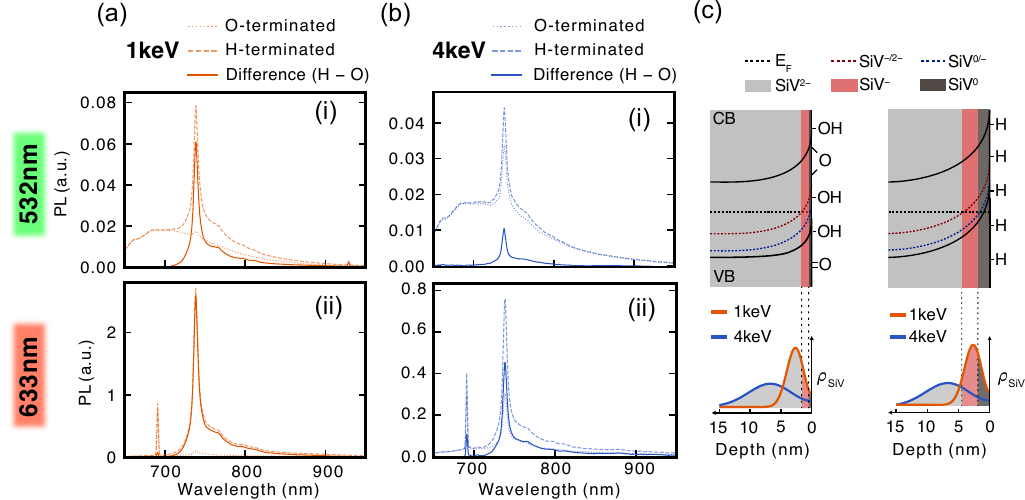}
    \caption{Surface doping effects on the silicon vacancy center using hydrogen and oxygen surface terminations at different ensemble depths. (a)-(b) demonstrates the effect of surface doping on the silicon vacancy fluorescence in a type IIa (< 5ppm nitrogen) substrate implanted with silicon vacancy at energies of 1keV and 4keV respectively. We compare the oxygen and hydrogen terminated states and show the relative change via a subtraction of the hydrogen terminated state by the oxygen terminated state in the case of illumination by 532nm (a)(i) and (b)(i), vs 633nm excitation (a)(ii) and (b)(ii). (c) is provided as a schematic from known charge state transition energies \cite{galiInitioStudySplit2013,garcia-arellanoPhotoInducedChargeState2024} to illustrate the relevant charge state transitions and how they're influenced by selecting for different depths and for alternate terminations.}
    \label{fig: hterm oterm}
\end{figure*} 

For this study, Type IIa $\langle 100 \rangle$ single crystal diamond samples with $<$ 5\,ppm of substitutional nitrogen (Element Six) were implanted with \textsuperscript{28}Si\textsuperscript{+} at a dose of $10^{13}$ions/cm$^2$ at an angle of 7\deg\, with energies of 1, 3, and 4\,keV (Coherent Ion Implantation). Although formation of SiV appeared post-implantation, additional SiV centers were created via annealing at 700-900\deg C (see supplementary information figure 1). Samples were hydrogen terminated by annealing in forming gas (5\% H\textsubscript{2} and 95\% N\textsubscript{2}, purity of 99.999\% (BOC)) for 4hrs at 700\deg C and later for 2hrs at 900\deg C. Oxygen terminations were restored via acid cleaning using Bristol boil (5\ce{H2SO4}:1\ce{NaNO3} at 500\,\deg C) and then piranha (4\ce{H2SO4}:1\ce{H2O2} at 120\,\deg C) for 15 minutes each. A Raman InVia Renishaw spectrometer with 532nm and 633nm excitation sources was used to excite and record diamond photoluminescence under hydrogen and oxygen terminations. For solution gated fluorescence measurements 532nm and 660nm was used to excite the sample in a conductive solution with the photoluminescence recorded using an Andor Neo sCMOS camera or a Princeton instruments Acton SP-2300i spectrometer with a grating of 300g/mm. For more details on the experiment and sample, see the supplementary information. 


\begin{figure*}[t]
    \includegraphics{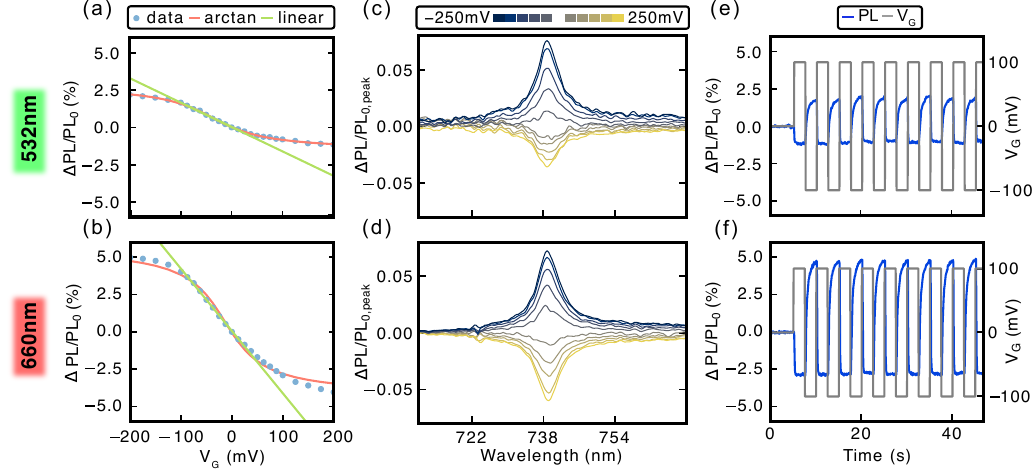}
    \caption{Dynamic control of \SiVm charge state populations by electrolytic solution gating. The top panel ((a),(c),(e))) represents explorations under 532nm laser while the bottom panel ((b),(d),(f)) duplicates the same analysis under 660nm illumination. (a) and (b) develops the extent of the charge gate fluorescence modulated between the range of -200mV and 200mV and fits an arctan across the full region and a linear fit across the center linear region (derived from the center of the arctan fit).(c) and (f) copies the treatment in (a) and(d) panel but reroutes the fluorescence to a spectrometer to verify the SiV fluorescence. Spectral normalization is done by dividing through every spectrum by the fitted \SiVm zero phonon line and then subtracting the spectrum at zero bias for comparison between the two illuminations.(e)(f) shows a square wave voltage switch using $\pm$ 100mV applied voltage and the corresponding fluorescence variation over time. }
    \label{fig: charge gating fig}
\end{figure*}

At room temperature both the 
\SiVmm and \SiVo are dark, leaving \SiVm as the only fluorescent charge state \figref{fig: figure1}(a)-(b). For near surface SiV populations oxygen terminations favor more electronegative charge states (\SiVmm,\SiVm) compared to hydrogen terminations due to upward band bending at the surface\cite{broadwaySpatialMappingBand2018}. Therefore, changes in \SiVm fluorescence incurred by moving from an oxygen terminated state to a hydrogen terminated state provides a method to determine the dominant charge state populations. 

Under 532nm illumination there is an increase in the \SiVm fluorescence shown when moving from the oxygen to the hydrogen terminated state in both the 1keV and 4keV (\figref{fig: hterm oterm}(a)(i) and (b)(i)). To show the contrast between the surface terminations and remove the unwanted background NV photoluminescence, the difference spectrum is shown in the solid line. The increase in \SiVm fluorescence suggests that the dominant charge state conversion process was between dark \SiVmm to \SiVm centers. However, exciting SiV centers with 532nm excitation is also known to introduce potential photo-ionization of SiV centres as well as nearby electroactive defects\cite{nicolasSubGHzLinewidthEnsembles2019,zhangNeutralSiliconVacancyCenters2023,dhomkarOnDemandGenerationNeutral2018}.  Therefore, spectra were also recorded under 633nm illumination which is known to be more weakly interacting with P1 and NV centers\cite{hausslerPhotoluminescenceExcitationSpectroscopy2017,woodWavelengthDependenceNitrogen2024}. When comparing the difference spectrum (solid line) between the two samples \figref{fig: hterm oterm}(a) vs (b) the change in fluorescence in the 1keV is approximately six times larger than the 4keV implantation for both illumination wavelengths. Given the implantation energies have a different baseline in their charge state populations the similarity in the scaling between the illumination wavelengths and the implantation energies suggests that the ionization of NV and P1 centers plays a minimal role in the reported \SiVm fluorescence for the excitation powers used. The process of moving between these hydrogen and oxygen terminated states can be repeated without loss of fluorescence, suggesting it is a reversible process as Zhang et al. \cite{zhangNeutralSiliconVacancy2023} showed for single SiV centers, and replicated for ensembles in supplementary information (Figure 2).

The increase in \SiVm fluorescence may be explained by the band bending profiles depicted in \figref{fig: hterm oterm}(c) which display the relative charge state populations under oxygen and hydrogen terminated surfaces. The charge transition energies are in alignment with literature and provide a depth dependent range for the different defect populations \cite{galiInitioStudySplit2013,thieringInitioMagnetoOpticalSpectrum2018,garcia-arellanoPhotoInducedChargeState2024}. The band-bending profile is overlaid with a Stopping Range In Matter (SRIM) simulation of silicon implanted into diamond to provide a guide for the SiV populations as a function of depth for each surface termination.

Interestingly, the oxygen terminated 1keV sample significantly suppresses the \SiVm charge state, while hydrogen terminating enhances the \SiVm population to a larger extent than the 4keV sample. These results confirm that the average depth and width of the implantation are key parameters in the optimization of SiV charge state populations but can be used alongside surface terminations to provide a means to stabilize desired charge states. However, surface termination only provides a coarse means to control the diamond surface charge and ensuing charge state populations. In the case of hydrogenated diamond surfaces, finer control can be realized by application of ionic or aqueous electrolytic gate voltages.


To investigate solution gating, a third hydrogen-terminated sample was implanted with \textsuperscript{28}Si\textsuperscript{+} at 3 keV and annealed at 700\deg C, placing the SiV ensemble at a depth commensurate with prior demonstrations of solution-gated NV center ensembles\cite{mccloskeyDiamondVoltageImaging2022}. A Ti/Pt electrical contact was fabricated on one corner of the diamond sample by electron-beam evaporation prior to thermal hydrogenation. For measurement, the sample was immersed in phosphate-buffered saline and electrically connected by contacting the Ti/Pt stack using a platinum wire electrode. Electrical control of the diamond was then probed via the injection and removal of holes through the Ti/Pt contact into the hydrogenated diamond 2-dimensional hole gas (2DHG).  

We first characterize the fluorescence response of the \SiVm ensemble to the application of DC solution gate voltages under both 532 nm (green) and 660 nm (red) excitation (\figref{fig: charge gating fig}(a)(b) respectively). A monotonic decrease in \SiVm emissions is observed as the potential of the diamond surface (with respect to the platinum bath electrode) is increased. We attribute this behavior to the conversion of \SiVm to \SiVo via capture of injected holes. As the increased \SiVm signal we previously observed when moving from oxygen to hydrogen termination indicates conversion from \SiVmm as a result of upward band-bending, we would anticipate that positive biases applied to the diamond surface in solution should further increase the \SiVm population. On the contrary, we observe decreases in \SiVm emission under positive biases, suggesting that hydrogen-termination alone is sufficient to deplete \SiVmm in our shallow ensembles. The fluorescence response to voltage resembles a broadened Fermi-Dirac function which we phenomenologically fit with an arctan function for quantification. The center point derived from these fits shifts by approximately +20mV going from green to red excitation schemes, which may result from surface photovoltage effects\cite{kronikSurfacePhotovoltageSpectroscopy2001,rezekHydrogenatedDiamondSurfaces2004} or changes in photoionization rates of \SiVm centers\cite{woodRoomTemperaturePhotochromismSilicon2023,garcia-arellanoPhotoInducedChargeState2024}. By fitting a linear region $\pm$50mV across the center point, we extract a small-signal fluorescence contrast of 0.0172 $\pm 0.0006$ \%/mV and 0.0418 $\pm 0.0008$ \%/mV for green and red illumination respectively. The applied potentials required to saturate the ensemble charge state switching effect here are more than an order of magnitude below those reported using solid state schemes \cite{brayElectricalExcitationChargestate2020,riegerMitigatingTransitionSiV$^$2025}. Furthermore, the linear response across $\pm$50mV about the curve center opens up the possibility of straightforward solution voltage sensing in the near infra-red similar to that shown for NV ensembles \cite{mccloskeyDiamondVoltageImaging2022}. 

\figref{fig: charge gating fig}(c)(d) verifies the spectral characteristics of the voltage dependent \SiVm fluorescence. Under 532nm illumination there is an NV background (Supplementary Figure 7), but by normalizing to the \SiVm zero phonon line (ZPL) at $V_G = 0$ and subtracting the measured spectrum from the \SiVm spectrum at $V_G = 0$, the relative change in \SiVm fluorescence is comparable under both illumination schemes. Under 532 nm excitation, asymmetry in the \SiVm spectral response across zero bias is observed, consistent with the observations in \figref{fig: charge gating fig}(a). The symmetry is preserved in \figref{fig: charge gating fig}(b) and (d) under 660nm excitation. The peak wavelength of the \SiVm emission was found to be redshifted which we attribute to the local strain near the surface of the diamond from the ion implantation\cite{angellUnravelingSourcesEmission2024}.

For dynamic charge state switching, we then apply a low power alternating gate voltage (square wave) of $\pm$ 100\,mV to demonstrate repeatable charge state switching of the SiV ensemble in \figref{fig: charge gating fig}(e) and (f). Over timescales of minutes, we did not observe any degradation of the switching due to the electrolytic solution gate. 
We observe response times on the order of hundreds of millisecond which contrasts with existing explorations of \SiVmm to \SiVm interconversion\cite{riegerFastOptoelectronicCharge2024} suggesting that response times here are not limited by the intrinsic electronic properties of the SiV defects\cite{riegerFastOptoelectronicCharge2024}. We primarily attribute the speed of the response to low surface conductivity due to background nitrogen and vacancies known to affect carrier density and mobility respectively\cite{risteinSurfaceConductivityNitrogendoped2002,zouProtonRadiationEffects2020}. To minimize the impact of defects such as vacancy/divacancy populations a slice of the 3keV sample went through a secondary anneal at 1100\deg C  \cite{tetienneSpinPropertiesDense2018,healeyProductionHighYield2026} and subsequently showed a 40\% reduction in the response times (Supplementary Information figure 5(c)-(f)). Further improvements could include using material with lower nitrogen populations to further increase surface conductivity. For the asymmetric response we observe only a small variation in time constants under red and green illumination, suggesting the dynamics of the time constant are not dominated by defect photoionization (Supplementary Information figure 5(a)-(d)). The asymmetric response we attribute to a Schottky barrier formed at the diamond-metal interface. Titanium carbide is known to provide an ohmic contact to diamond when annealing in vacuum\cite{jinguUltrashallowTiCSource2010}, however in hydrogen atmospheres, as done here, titanium hydride can form and may increase the barrier height at the surface\cite{depoverEffectTiCHydrogen2016,bootHydrogenTrappingEmbrittlement2024}. More optimized metallization strategies (for example annealing in vacuum conditions prior to implantation), together with increased contact surface areas \cite{riegerFastOptoelectronicCharge2024,riegerMitigatingTransitionSiV$^$2025} could further improve response times in future devices allowing the fundamental charge state switching limits to be achieved. Nonetheless, the temporal dynamics of the electrical charge state control of the \SiVm $\leftrightarrow$ \SiVo have not been demonstrated previously and as such we can provide an upper bound on the time constant of the switching speed as 220.2 $\pm$ 30.8 ms for \SiVo $\rightarrow$ \SiVm conversion and 65.5 $\pm$ 16.6 ms for \SiVm $\rightarrow$ \SiVo conversion.

This work has characterized the impacts of ion implantation depth, surface termination, and solution gate potentials on the charge state behaviors of shallow ensembles of implanted silicon-vacancy centers in diamond. 
Following ensemble formation by silicon ion implantation and vacuum annealing, we found that shallower ensembles formed by 1\,keV implantation exhibited increased \SiVm emissions when compared to a deeper ensembles formed by implantation at 4\,keV. Furthermore, we observed an increase in the relative population of the \SiVm charge state for both implantation energies upon switching from oxygen termination to hydrogen termination. Taken together, these results indicate that the Fermi level of our type IIa diamond samples lies above the SiV\textsuperscript{-/2-} adiabatic transition energy. 
Next, we measured the dependence of \SiVm fluorescence on an applied aqueous electrolytic gate bias and demonstrated a reversible and stable switch between the neutral and negative SiV center charge states. Fluorescence contrasts of 0.0172 $\pm 0.0006$ \%/mV and 0.0418 $\pm 0.0008$ \%/mV were reported when using 532\,nm and 660\,nm illumination respectively. This work demonstrates the first electrolytic gate based control of SiV centers in diamond which can be achieved with an order of magnitude reduction in applied voltage when compared with prior electrical charge state control methods.
The stability and reversibility of electrolytic control paves the way for low voltage control of shallow SiV centers for switchable photon sources; our results demonstrate charge state switching properties of ensembles, but future work could be extended to single shallow SiV defects in a similar vein to prior work focused on shallow NV centers \cite{grotzChargeStateManipulation2012}.
Low voltage solution gate control of SiV ensembles also illustrates their potential as a platform for charge state based voltage sensing and imaging in biology. Compared to previous work using NV ensembles \cite{mccloskeyDiamondVoltageImaging2022}, our demonstration of \SiVm charge state control uses red (660\,nm) excitation which could be further extended into the near-infrared region. This would lead to reduced phototoxicity on biological specimens of interest \cite{ichaPhototoxicityLiveFluorescence2017} while also reducing sources of background light generated by biological auto-fluorescence \cite{croceAutofluorescenceSpectroscopyImaging2014}. In investigating responses to liquid gate voltage, we confirmed via fluorescence spectroscopy that changes in intensity were restricted to the \SiVm charge state. In the context of voltage imaging, and in comparison to  NV centers, the substantially higher rates of \SiVm emission into the zero phonon line leads to a sharper emission profile that is much more easily multiplexed with common biological structure indicators (e.g., Green Fluorescent Protein). This proof of concept therefore opens up diamond voltage imaging to simultaneous recordings alongside other optical techniques.
While the results here present a first step towards high-fidelity control over shallow SiV charge states, a number of parameters merit further exploration. Future work will be needed to investigate the effects of varying the substitutional nitrogen density in the host material on the SiV ensemble charge states. For dynamic control, employing non-aqueous ionic liquids for solution gating would allow for higher applied voltages before the onset of oxidative damage to the hydrogenated diamond surface, potentially opening up a wider range of accessible charge states. Furthermore, all measurements performed here were undertaken at room temperature, which may explain our inability to observe \SiVo emissions at 945\,nm \cite{dhaenens-johanssonOpticalPropertiesNeutral2011}. Ionic liquid gating under cryogenic conditions could provide a pathway to stabilizing and interrogating populations of \SiVo and, potentially, \SiVp centers.

\vfill

\begin{acknowledgments}
C.P acknowledges funding from the University of Melbourne and the Graham Clarke Institute. D.J.M. acknowledges support through a University of Melbourne McKenzie Fellowship. This work was financially supported by the Australian Research Council (ARC)
Centre of Excellence for Quantum Biotechnology (CE230100021) and the ARC Discovery scheme (DP240102907). D.A.S. acknowledges support from the ARC Mid-Career Industry Fellowship (IM240100073). This work was performed in part at the Materials Characterisation and Fabrication Platform (MCFP) at the University of Melbourne and the Victorian Node of the Australian National Fabrication Facility (ANFF). A special thanks to Anders Barlow for his help in cryogenic experiments with the inVia Renishaw and Hunter Johnson for helpful discussions. 
\end{acknowledgments}

\bibliography{references} 
\end{document}